%% file: acc.tex
\documentclass[10pt,a4paper]{article}
\usepackage[hyperref]{acc}
\usepackage{times}
\usepackage{latexsym}
\usepackage{multicol}
\usepackage[sort, numbers]{natbib}
\usepackage{blindtext}
\usepackage{graphicx}

\usepackage[english]{babel}
\usepackage[autostyle, english = american]{csquotes}
\MakeOuterQuote{"}
\usepackage{url}
\usepackage{hyperref}

\usepackage{array}
\newcolumntype{L}[1]{>{\raggedright\let\newline\\\arraybackslash\hspace{0pt}}m{#1}}
\newcolumntype{C}[1]{>{\centering\let\newline\\\arraybackslash\hspace{0pt}}m{#1}}
\newcolumntype{R}[1]{>{\raggedleft\let\newline\\\arraybackslash\hspace{0pt}}m{#1}}
\usepackage{txfonts}
\usepackage{mathptmx}
\usepackage{color}
\usepackage{booktabs}
\newcolumntype{P}[1]{>{\centering\arraybackslash}p{#1}}

\usepackage{microtype}

\begin{document}

\twocolumn[
{{\huge A Study of Digital Appliances Accessibility for People with Visual Disabilities\par}\vspace{3ex}
    {\bf\large Hyunjin An\footnotemark[1], Hyundoug Kim\footnotemark[1], Seungwoo Hong\footnotemark[1], and Youngsun Shin\footnotemark[1]\par}\vspace{4ex}
    }

{\bfseries 
(Abstract) This research aims to find where visually impaired users find appliances hard to use and suggest guideline to solve this issue. 181 visually impaired users have been surveyed, and 12 visually impaired users have been selected based on disability cause and classification. In a home-like environment, we had participants perform tasks which were sorted using Hierarchical task analysis on six major home appliances. From this research we found out that home appliances sometimes only provide visual information which causes difficulty in sensory processing. Also, interfaces tactile/auditory feedbacks are the same making it hard for people to recognize which feature is processed. Blind users cannot see the provided information so they rely on long-term memory to use products. This research provides guideline for button, knob and remote control interface for visually impaired users. This information will be helpful for project planners, designers, and developers to create products which are accessible by visually impaired people. Some of the features will be applied to upcoming home appliance products.
\par}
\medbreak
\par\vspace{2ex}
]

\addtocounter{footnote}{+1}
\footnotetext{Samsung Electronics Co., Ltd}

\input{sections/01_intro}
\input{sections/02_relatedwork}
\input{sections/03_methodology}

\input{sections/04_results}
\input{sections/05_model}
\input{sections/06_implication}

\input{sections/07_conclusion}

\bibliographystyle{ieee_fullname}
\bibliography{acc}

\end{document}

%% file: sections/01_intro.tex
\section{Introduction}
Home Appliances are essential for everyone in everyday life. This of course cannot rule out visually-impaired. 
Visually impaired users' loss of vision and home appliances' lack of accessibility made using appliances a challenge. Lately, it came to a point where visually impaired users cannot use appliances without others' help. 

For computers, mobile devices and other screen-based products, various accessibility software, and guidelines~\cite{caldwell2008web} are available for visually impaired users. Different laws and acts~\cite{federal201021st,mclawhorn2001leveling,burgdorf1991americans} exist to enforce accessibility designs. Home appliances, however, tend to lack accessibility considerations and researches because at the moment it is pre-occupied taking care of its diverse product ranges which all has different control interfaces, and adding a feature directly impacts the products' cost.

This research aims to find where visually impaired users find appliances hard to use and suggest guideline to solve this issue.

181 visually impaired users have been surveyed, and 12 visually impaired users have been selected based on disability cause and classification. In a home-like environment, we had participants perform tasks which were sorted using Hierarchical Task Analysis (HTA)~\cite{stanton2006hierarchical} on six major home appliances (refrigerator, washer, air-conditioner, robot vacuum, microwave oven, and induction). After completing tasks, we conducted in-depth interviews to further evaluate visually impaired people's accessibility experience. 
From this research we found out that home appliances sometimes only provide visual information which causes difficulty in sensory processing for visually impaired users. Also, interface's tactile/auditory feedbacks were identical making it difficult for people to recognize which feature was processed. Blind users cannot see provided visual information so they had to rely on long-term memory to use products. 

This research provides guidelines for designing button, knob, and remote control interface for visually impaired users. This information will be helpful for project planners, designers, and developers to create products which are accessible by visually impaired people. Some of the features will be applied to upcoming home appliance products.

\begin{table*}[!t]
\centering
    \begin{tabular}{P{0.2cm}P{1cm}P{0.3cm}p{4.2cm}p{8.3cm}} 
        \toprule
        \textbf{ID} & \textbf{Gender} & \textbf{Age} & \textbf{Vision level} & \textbf{Frequently used home appliances}\\
        \midrule
        P1&	M&	47&	Blind, since 28 years old&	Fridge, Washer, Air Conditioner\\
        P2&	F&	38&	Blind, since 23 years old&	Fridge, Washer,
        Air Conditioner, Induction, Microwave, Vacuum cleaner\\
        P3&	F&	29&	Low vision, since birth&	Fridge, Washer,
        Air Conditioner, Oven, Microwave, Vacuum cleaner\\
        P4&	F&	28&	Blind, since birth&	Fridge, Washer, Dryer,
        Air Conditioner, Induction, Oven, Microwave, Vacuum cleaner, Air purifier\\
        P5&	F&	36&	Low vision, since birth&	Fridge, Washer,
        Air Conditioner, Oven, Microwave, Vacuum cleaner,
        Hood\\
        P6&	F&	49&	Blind, since 11 years old&	Fridge, Washer, Dryer,
        Air Conditioner, Induction, Oven, Microwave, Vacuum cleaner, Air purifier\\
        P7&	M&	35&	Blind, since 22 years old&	Fridge, Washer,
        Air Conditioner, Induction, Oven, Microwave, Vacuum cleaner, Air purifier\\
        P8&	F&	40&	Low vision, since birth&	Fridge, Washer,
        Air Conditioner, Induction, Microwave, Robot Vacuum cleaner, Air purifier\\
        P9&	F&	28&	Blind, since birth&	Fridge, Air Conditioner, Oven, Microwave, Vacuum cleaner\\
        P10&	M&	34&	Blind, since birth&	Fridge, Washer,
        Air Conditioner, Microwave, Vacuum cleaner\\
        P11&	M&	37&	Low vision, since 20 years old&	Fridge, Washer,
        Air Conditioner, Microwave, Vacuum cleaner, Air purifier\\
        P12&	F&	35&	Low vision, since 15 years old&	Fridge, Washer,
        Air Conditioner, Microwave, Vacuum cleaner, Air purifier\\
        \bottomrule
    \end{tabular}
\caption{Participants}
\label{tab:t1}
\end{table*}

%% file: sections/02_relatedwork.tex
\section{Related Work}

\subsection{Understanding visual impairment}
Most people who have a visual impairment are able to see something. This can vary from being able to distinguish between light and dark, to seeing large objects and shapes, to seeing everything but as a blur, or seeing a patchwork of blanks and defined areas~\cite{sardegna2002encyclopedia}. Visual impairment is a term used to describe all levels of sight loss. It covers moderate sight loss, severe sight loss and blindness~\cite{sardegna2002encyclopedia}.

In 2012, the World Health Organization estimated that of the 285 million visually impaired people in the world, 39 million were officially blind. In the U.S., the National Federation for the Blind estimates that around 6.6 million Americans are currently living with a visual disability~\cite{nfb}.
Also, the total number of legally blind students, ages 16 and up, enrolled in high schools in the U.S. is over 60,000~\cite{nfb}. 

Individuals who are visually impaired may be born with vision loss or develop a visual impairment later in life as a result of an accident or eye disease~\cite{afb}. Many different terms are used to describe the varying degrees of vision loss an individual may have~\cite{afb}. The terms "Visually impaired" and "Visual impairment" are used to include all individuals with decreased vision, regardless of the severity of vision loss or blindness. However, the following blindness terms and descriptions provide a better explanation of an individual's functional vision~\cite{afb}.

Visually Impaired: A person who is visually impaired has a decreased ability to see, even with corrective lenses, that adversely affects his visual access or interferes with processing visual information~\cite{afb}. The visual challenges an individual may have can range from not being able to see newspaper print to not being able to read print at all. Other challenges may include not being able to recognize a friend in a room until she is standing within arm's reach or until she identifies herself~\cite{afb}.

Blindness: The term "blindness" is typically used to describe individuals with no us-able vision or only the ability to perceive light~\cite{afb}.

Legally Blind: The term "legally blind" is a definition used to determine if individuals are eligible for government or other benefits as determined by the classification of legal blindness. Persons classified as legally blind have a central visual acuity of 20/200 or worse in the better eye with the best possible correction and/or a visual field of 20 degrees or less. For example, someone with an acuity of 20/200 may see some-thing at 20 feet the same as what someone with normal vision can see at 200 feet~\cite{afb}.

Low Vision: A person with normal vision typically has a visual acuity of 20/20 in both eyes and a visual field of approximately 160 to 180 degrees. An individual with low vision may have a visual acuity of 20/70 or worse and a visual field of 20 to 40 degrees or less. Individuals with low vision can often use optical devices, non optical devices, and environmental modifications to increase their visual functioning~\cite{afb}. 

Visual impairment / visual disability : A term that encompasses both those who are blind and those with low vision~\cite{corn2010perspectives}.
Visual impairment can also be classified into congenitally and adventitiously visual impairments~\cite{hupp2003cognitive}. The term congenital visual impairment refers to a condition of blindness or severe visual impairment that occurs at the moment of an absence. An individual whose blindness is prior to age 3 or 4 will in all probability not retain any visual imagery or visual memory, which provides important building blocks for the development of many basic and important concepts~\cite{lowenfeld1981berthold}.

Individuals who have been sighted but who have been lost have been described as having an adventitious visual impairment.

Individuals with congenital blindness often have different needs and experiences than do individuals with adventitious blindness. The effects of congenital blindness are often observed in such areas as sensory abilities, cognitive abilities, motor abilities, psychosocial abilities, travel skills, and independent living and vocational skills~\cite{moore1997foundations}.

\subsection{Studies on Digital appliance Accessibility}
People who are blind or visually impaired face severe difficulties using a growing array of everyday appliances because they are equipped with inaccessible electronic displays, often controlled with touch screens. 

Such appliances include microwave ovens, media players, digital blood pressure monitors, thermostats, and vending machines. Current technology to make printed documents accessible to persons with visual impairments is not equipped to handle the challenges of reading text on electronic displays, which is significantly different in appearance from printed text, and is often obscured by glare, reflections, or poor contrast~\cite{fusco2014using}.

A study by AFB (American Foundation for the Blind) focuses on blind user's home appliance usage to provide home appliance guideline designed for visually impaired~\cite{aaff}. To research accessibility of widely used home appliances, AFB divided control type into Accessible Controls, Inaccessible Controls, and Ambiguous Controls~\cite{aaff}. AFB found accessibility problems such as visually impaired users not being able to see the button or small display when setting the dish washer, defrosting food using Microwave and drying clothes using a dryer~\cite{aaff}.

A lot of the previous researches conducted on accessibility focus on computer and mobile. Or a high level of accessibility study was conducted such as problem discovery levels. This research will focus on home appliances and its various interfaces. And our research help researchers who design/develop/plan the industrial products.

%% file: sections/03_methodology.tex
\section{Methodology}

\begin{table*}[!t]
\centering
    \begin{tabular}{P{2cm}P{2.3cm}P{2cm}P{2cm}P{5.7cm}} 
        \toprule
        \textbf{Device} & \textbf{Model} & \textbf{Product Image}& \textbf{Interface type} & \textbf{Interface}\\
        \midrule
        
        Refrigerator	&4doors /RF85K9052XD	 &        \raisebox{-\totalheight}{\includegraphics[width=0.1\textwidth]{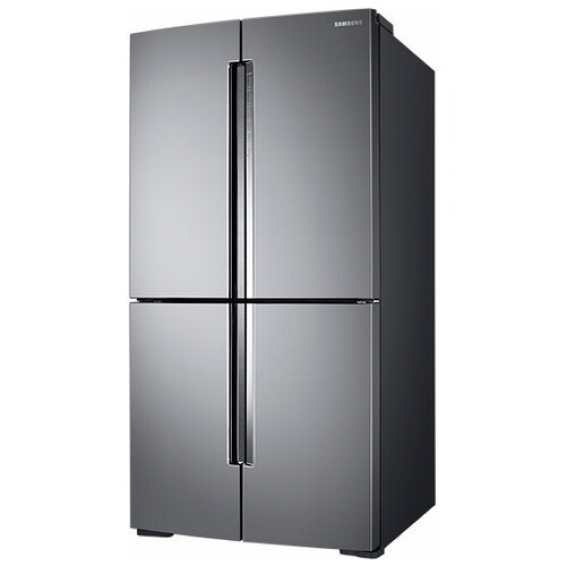}}&Touch button&        
        \raisebox{-\totalheight}{\includegraphics[width=0.35\textwidth]{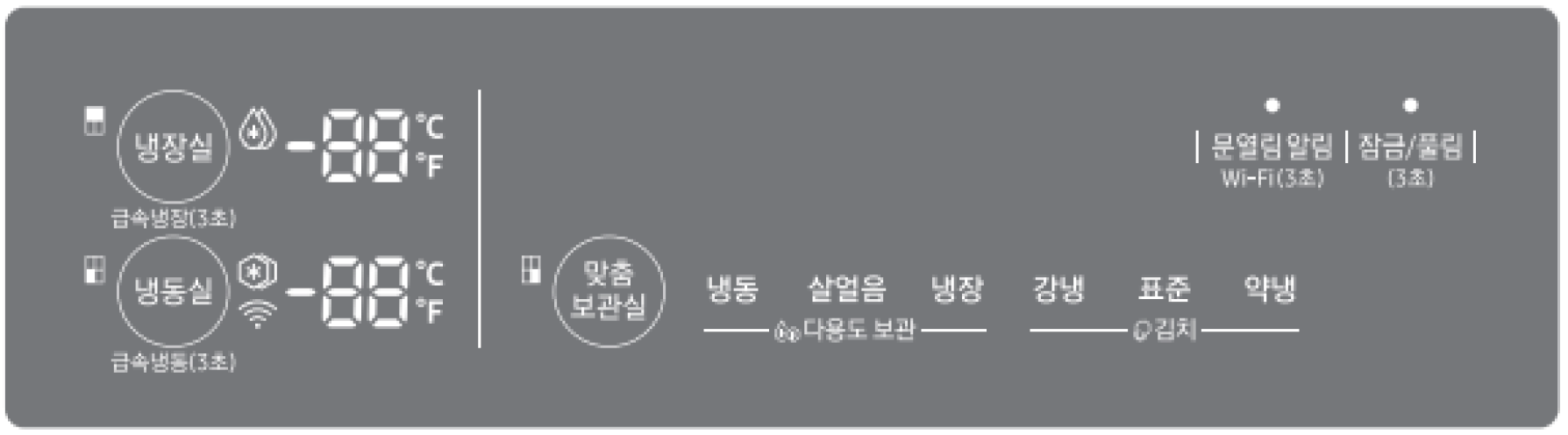}}\\
        
        Washer	&Front loader /WD23N9950KV	&\raisebox{-\totalheight}{\includegraphics[width=0.08\textwidth]{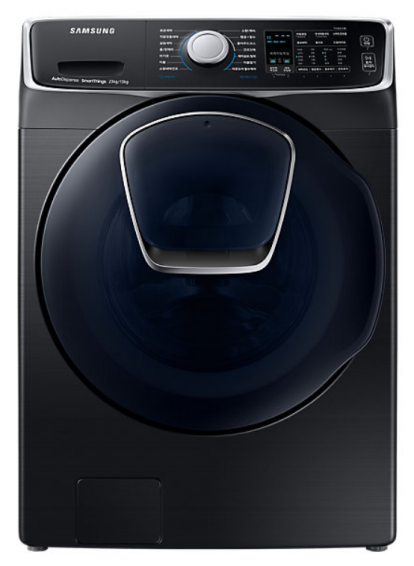}}&Knob, touch button&
        \raisebox{-\totalheight}{\includegraphics[width=0.35\textwidth]{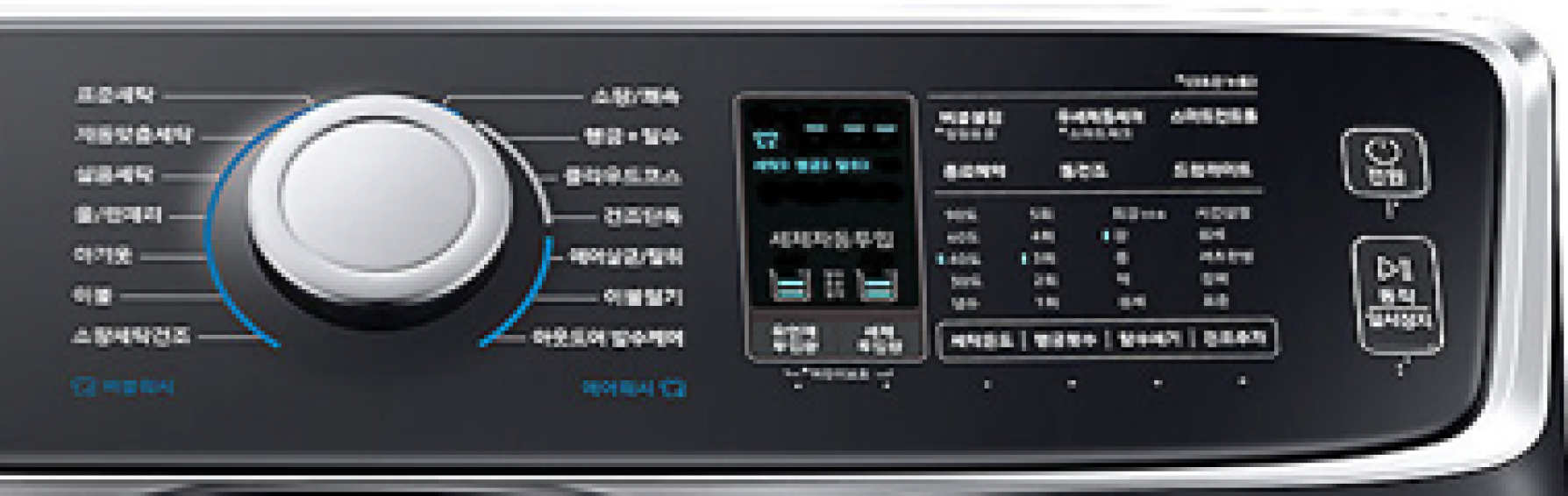}}\\
        
        Air conditioner	&Floor standing /Q9500	&\raisebox{-\totalheight}{\includegraphics[width=0.03\textwidth]{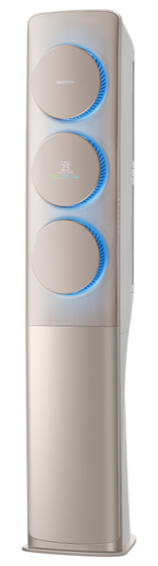}}&Tact button (RC)&
        \raisebox{-\totalheight}{\includegraphics[width=0.35\textwidth]{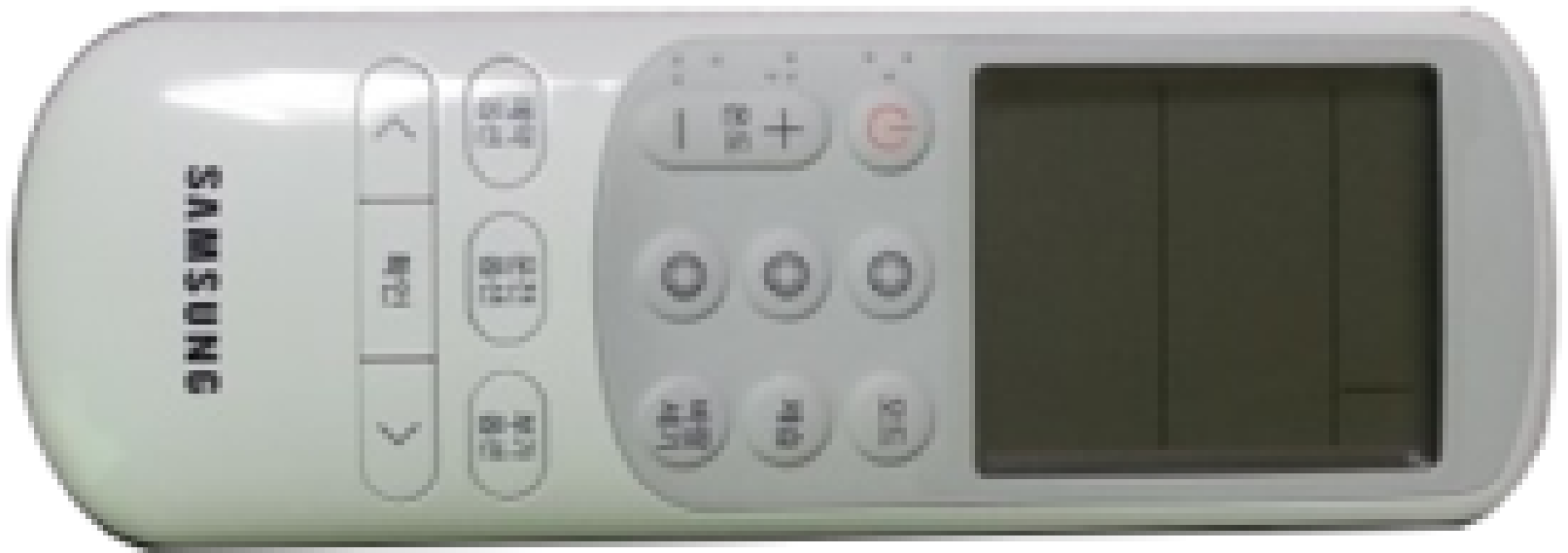}}\\
        
        Robot vacuum	&Robot Vacuum /VR7000	&\raisebox{-\totalheight}{\includegraphics[width=0.12\textwidth]{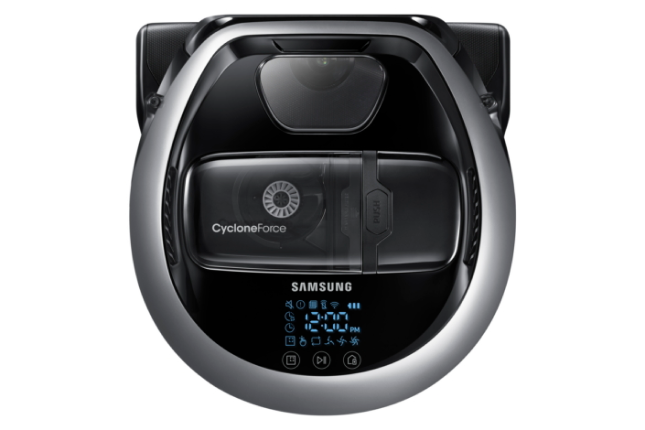}}&Tact button (RC)&
        \raisebox{-\totalheight}{\includegraphics[width=0.3\textwidth]{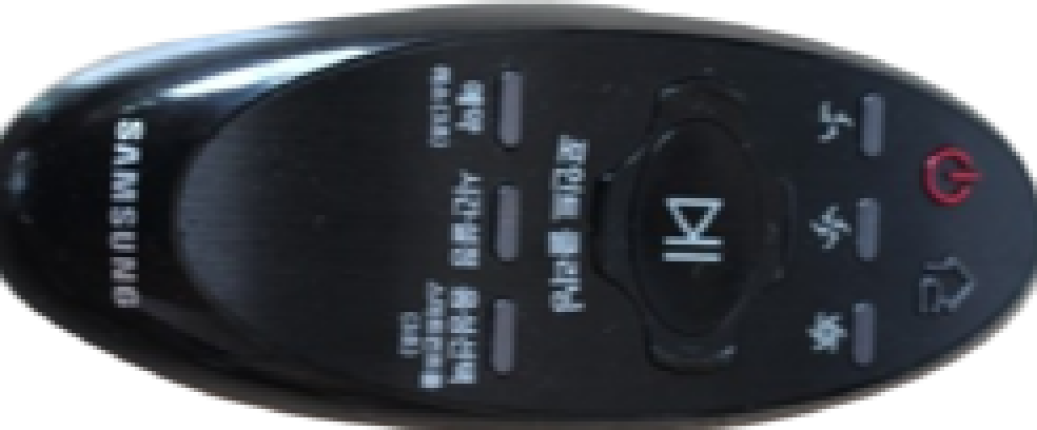}}\\
        
        Micro wave	&Gril-Solo /MW3500L	&\raisebox{-\totalheight}{\includegraphics[width=0.1\textwidth]{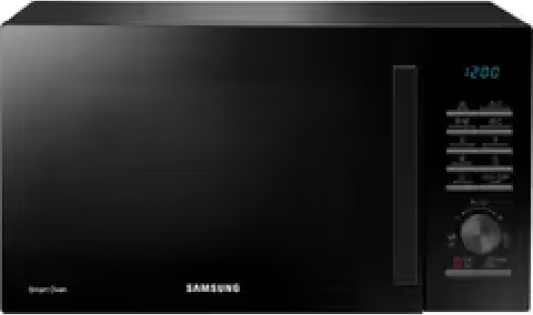}}&Knob, touch button&
        \raisebox{-\totalheight}{\includegraphics[width=0.2\textwidth]{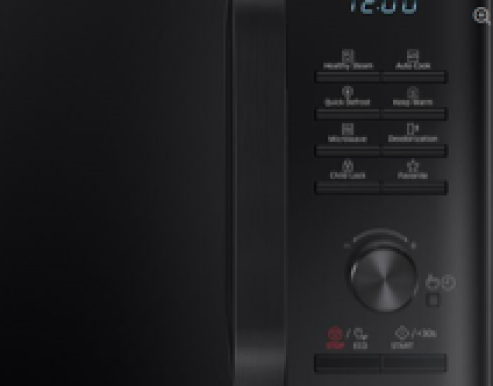}}\\
        
        Induction stove	&Built-in /NZ9000J	&\raisebox{-\totalheight}{\includegraphics[width=0.1\textwidth]{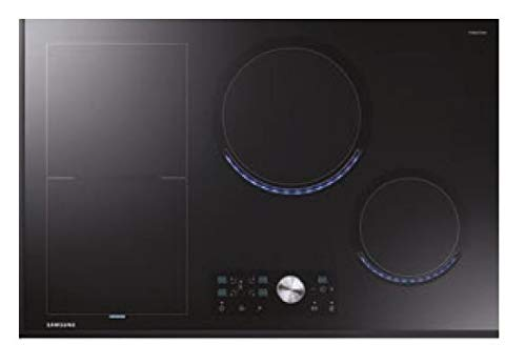}}&Slider, touch button&
        \raisebox{-\totalheight}{\includegraphics[width=0.2\textwidth]{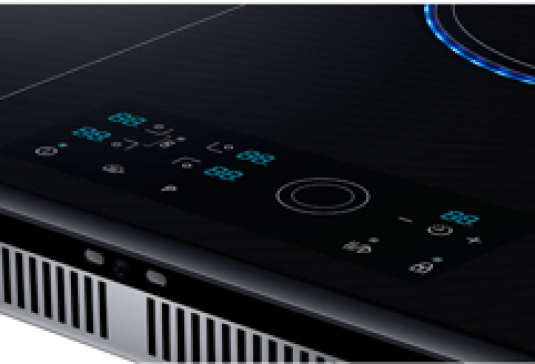}}\\
        \bottomrule
    \end{tabular}
\caption{Target devices}
\label{tab:t2}
\end{table*}


\subsection{Participant}
In order to find appropriate participants for the research, we conducted a pre-survey on 181 visually impaired users through visually impaired community "Nulbenmadang", and selected 12 participants based on the results. We recruited both blind and low vision participants based on medial classifications. Because there can be a difference in sensory, recognition, mobility, and language ability between innate and acquired blindness[12], we divided participants into three groups: 3 people with innate blindness, 4 people with acquired blindness and, 5 people with low vision. 8 participants were female and 4 were male. Participants' average age was 36.3 (SD = 6.69).  

\subsection{Target device}
Selecting appropriate appliances which can best represent that product was crucial as this research aims to research the current status of home appliances accessibility and present suggestions. 

Based on pre-survey, we selected 6 devices (refrigerator, washer, air conditioner, robot vacuum, microwave oven, and induction) which visually-impaired users mentioned were hard to use (table 2). We then selected common products with various interface controllers (e.g. touch button, tact button, dial, remote control). We picked average, everyday products to focus on home appliances' interface accessibility issue and feedback.

\subsection{Procedure}
As noted, the study was conducted in a home-like environment. Because participants were not familiar with test products, researchers explained each product until participants were able to use the products on their own. Tests began when participants were familiar with the product. 

We adopted an everyday use case scenario for the selected products and further categorized each task  using the Hierarchical Task Analysis method ~\cite{stanton2006hierarchical}. We tested products in different orders to avoid side effects due to order. Visually impaired participants conducted tasks for each product, and we analyzed whether the products were accessible or not. The entire procedure was video-recorded and interviews were conducted when participants failed a task or found the task difficult.

%% file: sections/04_results.tex
\begin{table*}
  \caption{Test and analyze 12 key tasks of household appliances for Blind and Low vision users.}
  \includegraphics[width=\linewidth]{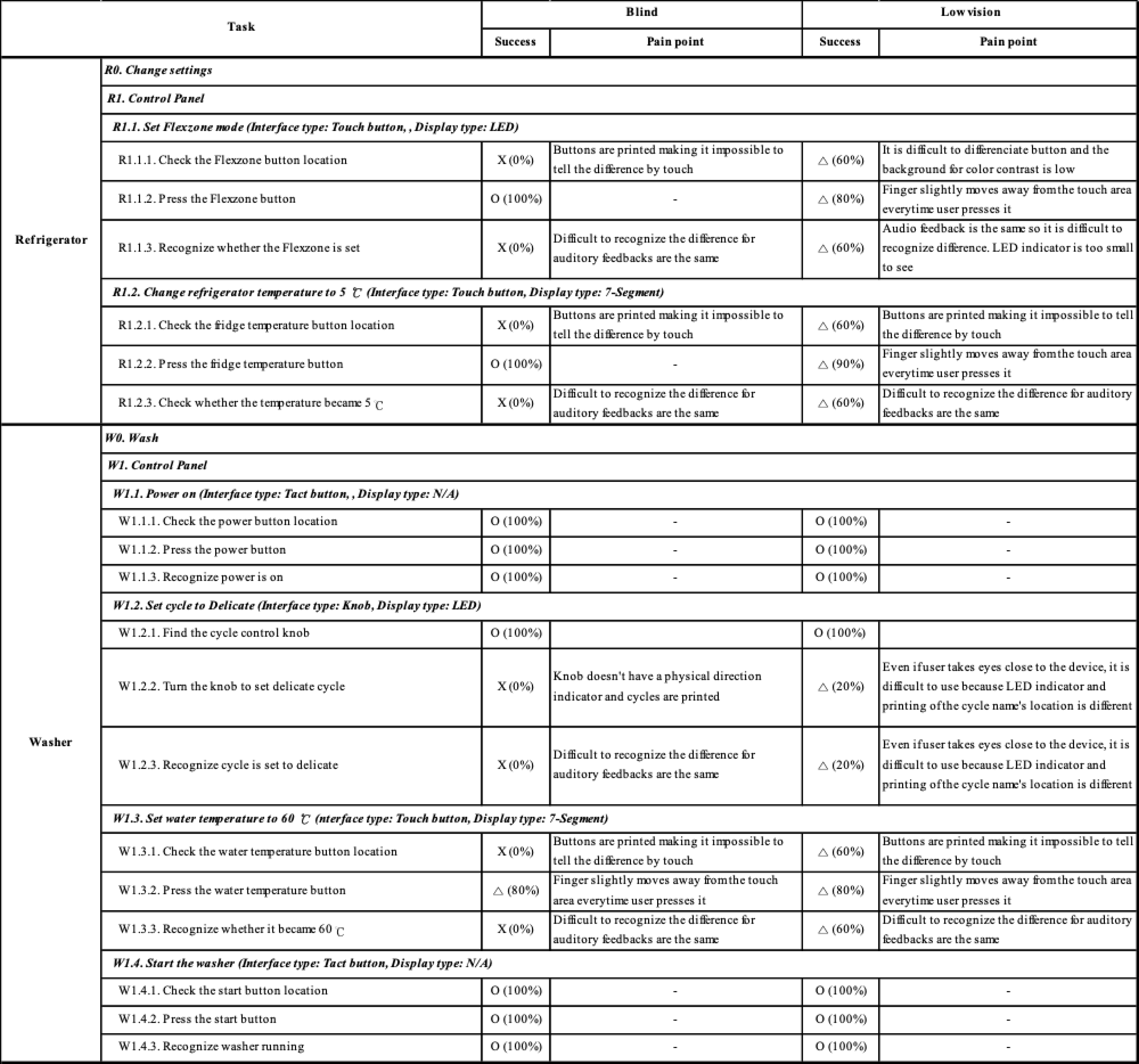}
  \label{tbl:excel-table}
\end{table*}

\begin{table*}
  \includegraphics[width=\linewidth]{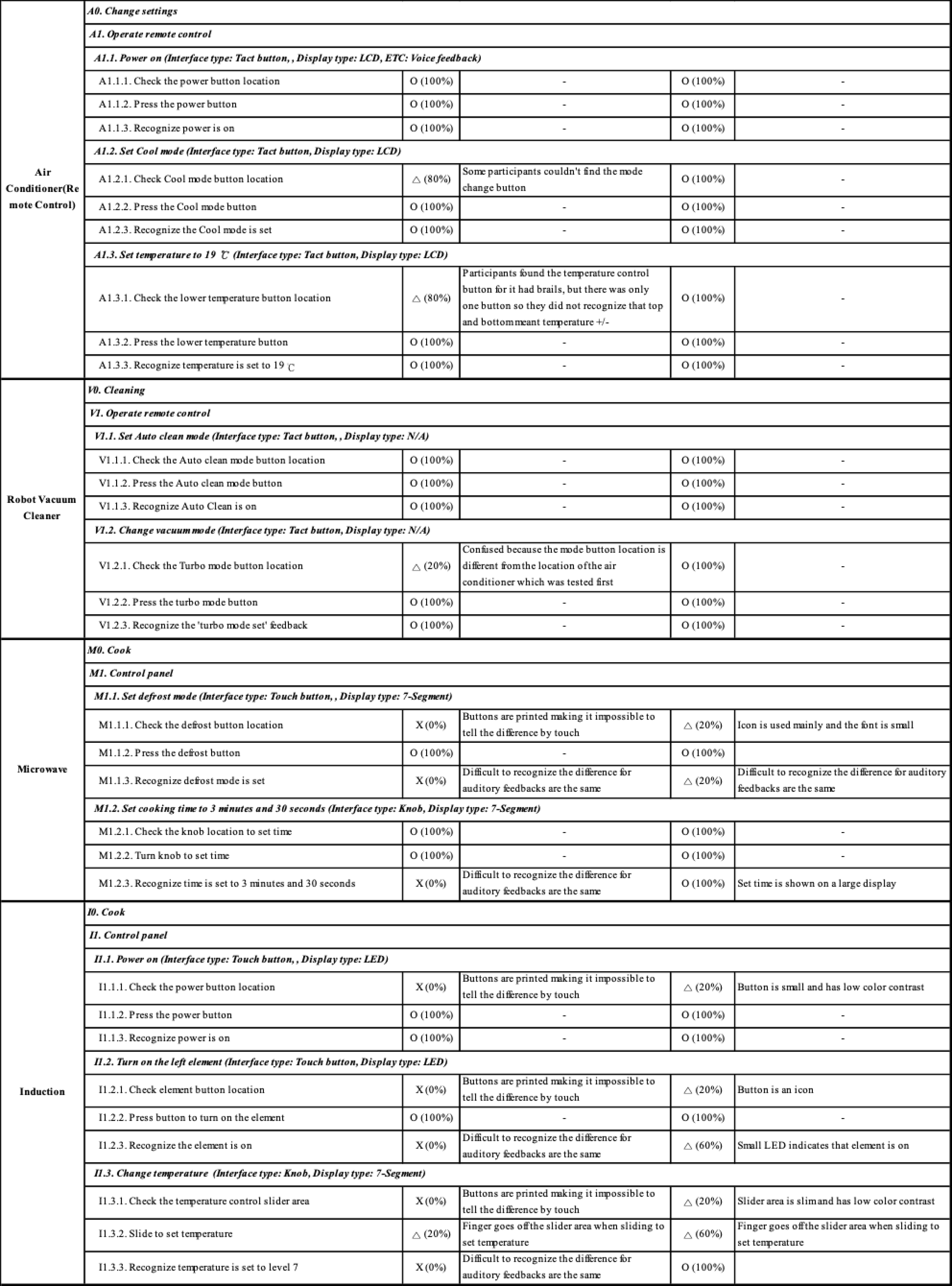}
  \label{tbl:excel-table2}
\end{table*}

\section{Results}
\subsection{Blind}
\subsubsection{Finding interface}
Blind participants had problems with finding the interface they wanted to control. Since it is impossible for them to see, blind participants used tactile senses to find buttons. Touch interface buttons, however, cannot be distinguished by touching, making it hard for participants to find them (R1.1.1, R1.2.1, W1.3.1, I1.1.1, I1.2.1, and I1.3.1 from Table 3). Due to the nature of touch buttons, accidental brushing of the buttons resulted in unwanted action taking place. 

\textit{"(sigh) This (touch interface) completely excludes visually impaired users for it was made for the users who can see. If it had a slight indent, it can be used as a reference, but this does not even have that"} (P2) 

Products with tact buttons (W1.1.1, W1.4.1, A1.1.1, A1.2.1, A1.3.1, V1.1.1, and V1.2.1 from Table 3), 
however, does not operate by brushing, so if users remember the exact location of the button, they had no problem with 
finding the button they wanted to use. 
Knob interfaces usually protrude so it was easier than buttons for blind participants to find.

\begin{figure}[!h]
    \centering
    \includegraphics[width=\linewidth]{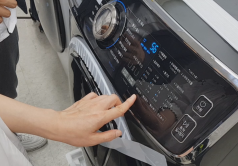}
    \caption{Blind participating using the touch button(washer)}
    \label{fig:f1}
\end{figure}

\subsubsection{Control}
Blind participants had no problem with motor skill, they were able to press buttons or control knob interfaces with little or no difficulty. 
When having to press touch buttons multiple times (R1.1.2, and W1.3.2 from Table 3), however, every time the button was pressed, finger moved away from the button (Fig. \ref{fig:f1}). If touch buttons have narrow spacing, finger moving lead to pressing the next button. 

Similarly, with the changing temperature using the touch slider task (I1.3 from Table 3), when participants found the slider, they had trouble with moving their fingers in line with the slider. both cases occurred because there was no method to help them correct their location tactilely.

\subsubsection{Recognition}
Most of the participants had trouble with recognizing whether tasks were correctly applied or not. Because they cannot confirm with vision, participants relied on their auditory senses. Most home appliances in the market, however, provide a same buzzer sound for feedback which makes it hard to figure out the status without viewing (R1.1.3, R1.2.3, W1.2.3, W1.3.3, M1.1.3, M1.2.3, I1.2.3, and I1.3.3 from Table 3). 

For example, with the setting the washer water temperature to 60℃ task (W1.3 from Table 3), blind participants had a hard time figuring out whether the temperature was going up or down, and what the set temperature was because the identical ''beeping” sound was given even if they found the correct temperature control button to press. 

If buzzer (part which makes the beeping sound) had a specific sound, however, the participants could recognize the action based on auditory feedback. For example, with the induction power on task (I.1.1 from Table 3), blind participants could tell power on and off apart when the power button was pressed, because musical scale going up auditory feedback was used when power turned on and musical scale going down was used when power turned off. 

Blind participants instinctively and definitely understood the feedback when auditory feedback was provided (A1.1.3, A1.2.3, A1.3.3, V1.1.3, and V1.2.3 from Table 3). For example, with the setting the cool mode using the air conditioner remote control task (A1.2 from Table 3), when the blind participants pressed the mode button, audio said ‘cool mode'. Through this auditory feedback, participants knew cool mode was set.

\textit{''I want this air conditioner. If there was a product like this (when I bought an air conditioner) I would have bought it. Voice control brings it to another level.''} (P7)

Voice feedback can provide clearer and intuitive feedback than products which use buzzer.

\begin{figure}[!h]
    \centering
    \includegraphics[width=\linewidth]{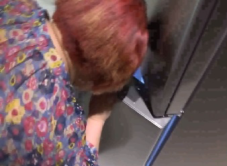}
    \caption{A low vision participant taking her eyes close to the button to see (refrigerator)}
    \label{fig:f2}
\end{figure}

\begin{figure}[!h]
    \centering
    \includegraphics[width=\linewidth]{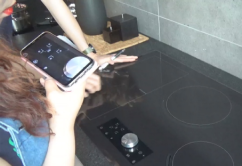}
    \caption{A low vision participant taking a picture of the interface with a smartphone and enlarging it to check (induction)}
    \label{fig:f3}
\end{figure}

\subsection{Low vision}
\subsubsection{Finding interface}
Participants with low vision have some sight left, so it was easier for them than the blind Participants to find interfaces. 
It was easier to identify an interface if it had high color contrasts between the background and the information, and used large icons and texts. 

Low vision participants who engaged in this research all had their own methods to make up for their low vision.  

First was to look closely (Fig. \ref{fig:f2}). Since it was impossible to distinguish an interface from a far, participants took their eyes close to it. This was the most frequently used method but refrained from using if an interface was located far or if it was not safe to do so (I.1,I1.2,and I1,3 from Table 3). People with low vision should be careful when using this method for it is not good for the eyes to view bright LED closely. 

Second was to use a camera (Fig.\ref{fig:f3}). Participants used smartphones or camcorders if it was hard or dangerous to take their eyes close to the products. Using a camera, participants enlarged images they wanted to view closely. Nowadays cameras support high definition images, so when participants enlarged the images they still kept their quality.

\subsubsection{Control}
Low vision participants have no problem with motor skill, they were able to press buttons or control knob interfaces with little or no difficulty.

Because low vision participants have some sight left, finger location changing when pressing the same button multiple times issue (R1.1.2, and W1.3.2 from Table 3) which occurred with the blind participants did not occur. For example, with the controlling temperature using the induction task (I1.3 from Table 3), participants took pictures of the interface with their smartphones and enlarged it to search wanted buttons (Fig 3.)

\subsubsection{Recognition}
Participants with low vision, like the blind participants, relied heavily on auditory feedback to recognize whether action set was applied correctly. They were able to understand information clearly and instinctively when appliances had a distinctive sound or provided a voice feedback.   

The main difference with the blind and the low vision participants was that the low vision participants can check visual feedback with some sight they have left. However, they face the same issue as finding interface, so they took their eyes close to the interface or used a camera to check whether the wanted action was applied correctly or not.  

Clear auditory feedback is necessary for both the blind and the low vision users. 

%% file: sections/05_model.tex
\section{Interpretation from Human-Machine Interface point of view}
\begin{figure}[!ht]
    \centering
    \includegraphics[width=\linewidth]{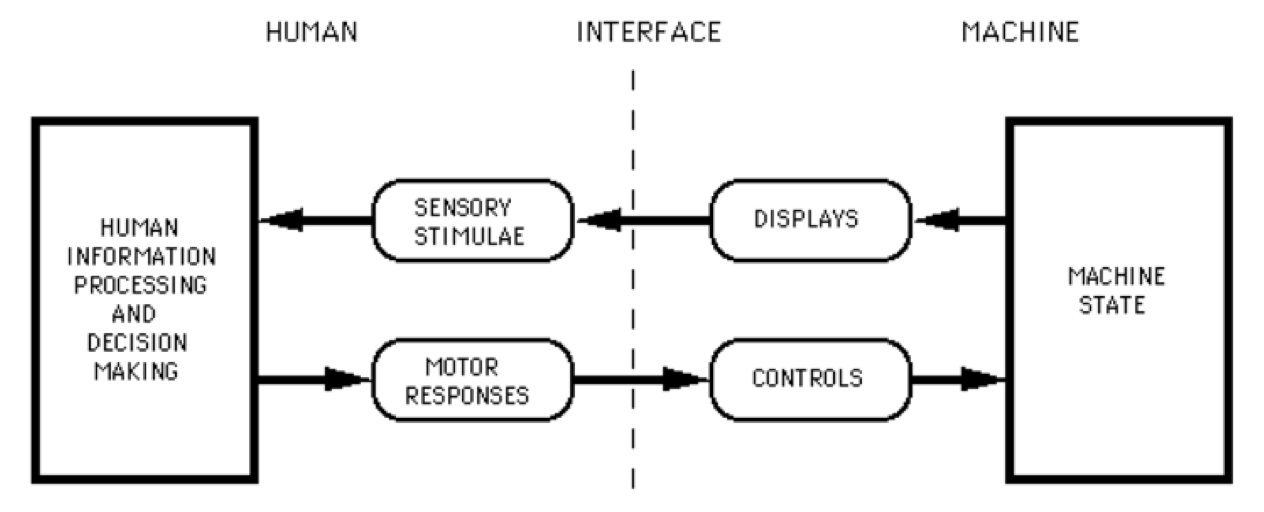}
    \caption{MacKenzie's The human-machine interface model~\cite{mackenzie1995input}}
    \label{fig:f4}
\end{figure}

MacKenzie, like figure 4, simplifies the human and machine to three components~\cite{mackenzie1995input}. The internal states of each interact in a closed-loop system through controls and displays (the machine interface) and motor-sensory behavior (the human interface)~\cite{mackenzie1995input}. This section aims to divide interaction between users and home appliances using MacKenzie's human-machine interface model (Fig. \ref{fig:f4}) to specify existing problems. 

First is to look into display~ sensory stimulate process where user and machine's initial interaction occurs.

\subsection{Displays ~ Sensory Stimulate}
Appliance (machine) usually represents its idle state before the first interaction with user occurs. Being in an "off" mode also represents one of many visual state representations. 

People without disability can take in this visual state representation using vision. People with visual impairments, however, cannot take in state representations if status is displayed only using visuals. A method to check information using senses other than vision, or information shown using a combination of two or more senses need to be provided.
\subsection{Sensory Stimulate ~ Human Information Processing and Decision Making}
Acquired senses process information given to make decision. With the home appliances used in the research, which are everyday products, all the information is shown live visually, but not auditory and tactilely. Users without disability can rely on short term memory (STM) to use appliances, but visually impaired users have to rely on long term memory (LTM). Visually impaired users have to educate themselves on appliances usage in order to use them without viewing.

For example, with the changing wash cycle course to delicate task (W1.2 from Table 3), people without disability can figure out what the entire washing cycles are by looking, remember them momentarily, and immediately make a decision that delicate is the right course to use. They can also use visual senses to figure out where the delicate cycle is located. 

Visually impaired users, however, cannot figure out the entire washing cycles the washer provided visually. They have to learn in advance the existence of delicate cycle and its location. Visually impaired users have to remember features and interface location of that feature in order to use appliances. A method to assist them to learn or to get feature information easily is needed. 
\subsection{Human Information Processing and Decision Making ~ Controls}
Visually impaired users do not have difficulty with controlling devices after decision has been made for they do not have problems with motor skill.
\subsection{Controls ~ Sensory Stimulate}
Controls and sensory stimulate component of the model corresponds to feedback on action. If only visual feedbacks are provided, visually impaired people who cannot accommodate visual senses will not be able to process information. Therefore, combined multi modal feedback is required. 
This multi modal feedback, however, cannot be identical if it has one interface with many states. 

For example, setting the washer temperature to 60℃ task (W1.3 from Table 3), when pressing a temperature button, it changes from Cold-30℃-40℃-60℃-90℃-Cold. For every change, same beep sound is repeated making it impossible to tell what the set temperature is without viewing. 

However, with the turning air conditioner power on task (A1.1 from Table 3), different sound is used when power is on and off, which makes it easy to figure out its state.  

When controlling using one interface and that interface has more than one state, at least one reference point feedback should be provided to assist in figuring out different states.

%% file: sections/06_implication.tex
\section{Appliance design implication}
\subsection{Touch/Tact Button}
\subsubsection{Provide a method to differentiate buttons}
Touch buttons do not have tactile but only visual presence. Touch buttons easily activates when lightly touched making it difficult for visually-impaired users to distinguish the buttons. Tactile information near touch button should be provided. 

Usually tactile dots are provided on top or bottom of the buttons. 
If tact buttons have different size and shape for every button, and that is enough to tell the difference, then tactile information may not be necessary. 
\subsubsection{Provide auditory feedback}
Auditory feedback is an effective method for visually impaired users to recognize button usage. It lets users to figure out devices' status and whether it activated or not almost immediately after using. Representative auditory feedbacks are voice feedback and beep sound. 

Voice feedback provides direct and accurate information, therefore is the best solution for improving accessibility for visually impaired users. It, however, increases the overall cost of the product with its expensive parts. Beep sound, on the other hand, is cheaper to produce but lacks clarity in information for it's hard to provide exact information using melody.

Since touch button do not have tactile feedback when pressed, auditory feedback must be provided (Mobile uses haptic, but haptic feedbacks are rare with appliances for they have bigger interfaces). Like mentioned in section 5.4, if one button has more than two functions, providing different auditory feedback for each function will help visually impaired users to distinguish its status (e.g. tell on/off sounds apart)

\subsubsection{Consider spacing between buttons}
Touch button is a very sensitive interface. There are high percentages of people touching (selecting) buttons accidently resulting in unwanted actions. Wider spacing between buttons is necessary to prevent wrong touches. 

Tact buttons, on the other side, compared to touch buttons, does not require wide spacing between buttons because users can touch the interface and it would not result in actions unless pressed. When a lot of buttons need to be squeezed into a small area like the remote control, using a tact button is favorable.
\subsection{Knob(dial)}
\subsubsection{Tactile direction indicator}
All of the knobs used in this research did not have a tactile marking. It was difficult for visually impaired users to figure out which direction the knob was facing. Some of the participants wanted tactile markings for knobs. 

\textit{''Washer at home has a round sign on top of the knob which indicates direction its pointing at, but this does not [have an indicator]'' }(P4)

People with visual impairments will be able to recognize the direction the knob is indicating, if there is a tactile indicator at the direction it is pointing.
\subsubsection{Knob notches}
Both the washer and the MWO used in the research had a knob, but they provided different tactile experiences. When turned, the washer knob notched whereas the MWO had little to no tactile differences. P2, a visually impaired participant, mentioned that this has a huge difference in accessibility.

\textit{''The bigger the notching feel, the better. You can tell how many times this (knob) turned instantly...''} (P2)

Participants were more careful and sensitive when using a MWO knob than a washer knob. People with visual impairment are more likely to accurately recognize the difference if there is a bigger notching feel when turning a knob.
\subsubsection{A knob which does not turn infinitely}
One of the characteristics of all of the appliances with a knob used in this research was that the knob turned infinitely. Blind participants P2, P4, P6, P7, and P10, expressed a strong dislike towards this infinite turn. 

\textit{''We can use it as a reference point if this [the knob] did not turn infinitely but stopped at one end.''} (P10)

If a knob stopped at one end instead of turning infinitely, visually impaired users can use it as a physical reference point; especially when a knob deals with numbers with order, like going from a smaller value to a larger value (e.g. setting time task with MWO, M12 from Table 3). A knob which does not turn infinitely will bring a huge improvement in home appliances accessibility. 
\subsubsection{Provide auditory feedback}
Like mentioned in section 6.1, auditory feedback can be a strong increase in accessibility with a knob as well. A knob is a representative input method which has one interface with many functions. Auditory feedback is required in order to tell different status apart. Methods such as change in volume, like when changing a radio volume, and difference in pitch can be used.
\subsection{Touch Slider}
\subsubsection{Provide tactile guidance}
All of the blind participants could not use the induction temperature control slider for it was a touch interface. Because it was a touch interface, the participants could not find the interface and even if they did, with no tactile guidance, it was difficult to drag touch slider in a straight line. Even when told the starting point, a lot of the participants' finger went off the slider area. Tactile indicator to distinguish starting and end point, and physical guidance to help clearly move through the slider area are needed.
\subsection{Physical Aspects}
\subsubsection{Place interface at an area which is not high or deep}
The biggest difference between mobile devices, computer, and home appliance accessibility is that home appliances should consider physical aspects more than the other devices. There are a lot of physical aspects, but when compared to mobile devices or computer; physical input, height, and depth are the most distinct. Out of the three, height and depth are the more relevant when it comes to visually impaired accessibility. Low vision participants take their eyes close to the interface to see well.

This, however, is impossible if the interface is located too high or too deep. P12, a low vision participant, who took part in the research, mentioned that it is difficult to use a top loader washer at home because the interface is located at the deep back of the washer. It is impossible to take ones' eyes close to the interface. For a similar reason, P5 picked hood as the most difficult home appliance to use. Hood interface is located too high to take one's eyes close to. Product's interface should be located not too high and not too deep. This applies to elderly without disability and children as well. 
\subsubsection{Interface location which follows set rules}
Blind participants all succeeded in turning the power on using the air conditioner remote control (A1.1 from Table 3). They succeeded because ‘the power button is located at the top left hand corner which is the same location as the remote control used at home'. Also they easily predicted that pressing top button will increase the temperature and down button will decrease the temperature when completing the air conditioner temperature control task (A1.3 from Table 3). Interface location which follows the society's norm increase visually impaired users' accessibility.

%% file: sections/07_conclusion.tex
\section{Conclusion and discussion}
This research aims to find and solve problems visually-impaired users have when using home appliance interfaces. A total of 12 blind and low vision participants conducted representative tasks using a total of 6 home appliances (refrigerator, washer, air conditioner, robot vacuum, microwave oven, and induction) in a home like environment.  
Through task evaluation, accessibility improvement points for visually impaired using tact/touch button, knob, touch slider, and physical aspects were found.

With Tact/Touch button, methods to tell buttons apart using size, shape, and protrusion needs to be provided. Also to help recognize button usage, auditory feedback should be implemented. With the touch button, to prevent false action, wide spacing is needed. With the knob, tactile marking should be present to assist figuring out which direction the knob is pointing at. Implementation of notches and non-infinite turn will better its accessibility and usage. Touch slides need tactile guidance so that users can figure out where the starting and ending point is, and drag without getting off of touch area. Lastly, as for appliances' physical aspects, height and depth should be considered when placing an interface. 

It will be beneficial if interface is designed following rules which are universally accepted. 

Solutions suggested per interface all mention that senses other than vision should be used when presenting information. Auditory and tactile feedback should be presented along with visual feedback for olfactory and taste interfaces are not yet available for home appliances. Also, reference point is a must because if identical audio and identical tactile marks are used, without a reference point, users cannot tell the difference apart. Having at least one reference point will greatly improve products' accessibility. 
 
This research analyzes what kind of problems visually impaired users face when using home appliances, and suggests specific methods per interface to solve these problems. Home appliances developers, designers, and product managers will be able to make more accessible products for not only visually impaired users, but also users without disability in terms of universal design, by following this research's design methods. We look forward to the world wide improvement of visually impaired user's accessibility.